\documentstyle [12pt]{article}
\title{The spectral properties of the Falicov-Kimball model in the 
weak-coupling limit}
\author{Pavol Farka\v sovsk\'y\\
Institute  of  Experimental  Physics,  Slovak   Academy   of
Sciences\\
Watsonova 47, 043 53 Ko\v {s}ice, Slovakia}
\date{}
\begin{document}
\baselineskip=20pt
\maketitle

\begin{abstract}
The $f$ and $d$ electron density of states of the one-dimensional 
Falicov-Kimball model are studied in the weak-coupling limit by exact 
diagonalization calculations. The resultant behaviors are used to examine 
the $d$-electron gap ($\Delta_{d}$), the $f$-electron gap ($\Delta_{f}$),
and the $fd$-electron gap ($\Delta_{fd}$) as functions of the 
$f$-level energy $E_f$ and hybridization $V$. It is shown that the 
spinless Falicov-Kimball model behaves fully differently for 
zero and finite hybridization between $f$ and $d$ states.
At zero hybridization the energy gaps do not coincide 
($\Delta_{d}\neq \Delta_{f} \neq \Delta_{fd}$),
and the activation gap $\Delta_{fd}$ vanishes discontinuously
at some critical value of the $f$-level energy $E_{fc}$. 
On the other hand, at finite hybridization all energy gaps
coincide and vanish continuously at the insulator-metal transition 
point $E_f=E_{fc}$. The importance of these results for a description 
of real materials is discussed.

\end{abstract}

\newpage

\section{Introduction}
The Falicov-Kimball model~(FKM) has become, since its introduction in 1969,
one of the most popular examples of a system of interacting electrons
with short-range interactions. The model was originally proposed to describe 
metal-insulator transitions~\cite{Falicov} and has since
been investigated in connection with a variety of problems such
as binary alloys~\cite{Freericks_Falicov}, the formation of 
ionic crystals~\cite{Gruber/Lebowitz}, and ordering in mixed-valence 
systems~\cite{Ramirez}. It is the latter language we shall use here, 
considering a system of localized $f$ electrons and itinerant $d$ 
electrons coupled via the local interaction $U$ and hybridization $V$. 
The Hamiltonian of the spinless FKM is 

\begin{equation}
H=\sum_{ij}t_{ij}d^+_id_j+U\sum_if^+_if_id^+_id_i+E_f\sum_if^+_if_i+
V\sum_id^+_if_i+h.c.,
\end{equation}
where $f^+_i$, $f_i$ are the creation and annihilation
operators  for an electron in  the localized state at
lattice site $i$ with binding energy $E_f$ and $d^+_i$,
$d_i$ are the creation and annihilation operators
of the itinerant spinless electrons in the $d$-band
Wannier state at site $i$.

The first term of (1) is the kinetic energy corresponding to
quantum-mechanical hopping of the itinerant $d$ electrons
between sites $i$ and $j$. These intersite hopping
transitions are described by the matrix  elements $t_{ij}$,
which are $-t$ if $i$ and $j$ are the nearest neighbors and
zero otherwise (in the following all parameters are measured
in units of $t$). The second term represents the on-site
Coulomb interaction between the $d$-band electrons with density
$n_d=\frac{1}{L}\sum_id^+_id_i$ and the localized
$f$ electrons with density $n_f=\frac{1}{L}\sum_if^+_if_i$,
where $L$ is the number of lattice sites. The third  term stands
for the localized $f$ electrons whose sharp energy level is $E_f$.
The last term represents the hybridization between the itinerant
and localized states.

In this paper we perform exhaustive numerical studies of the spectral
properties of the FKM with and without hybridization. While the static 
properties of the FKM are well understood at present~\cite{prb1,prb2} 
(including the picture of valence transitions), the dynamical properties 
of the model are still unclear. Even, the spectral properties of the $f$ 
electrons are not understood satisfactorily nor for $V=0$, where only a few 
exact results are known for the infinite-dimensional systems~\cite{Brandt,Zlatic}. 
No exact results are known for nonzero hybridization and $T=0$, with the 
exception of numerical results obtained on very small clusters~\cite{Park}. 
The first systematic study of dynamical properties of the FKM with 
hybridization has been performed recently by Craco~\cite{Craco} within 
so called static approximation. He studied the model in the strong-coupling
limit (large $U$) and found that the system is insulating for small values of 
hybridization (the size of the $f$ and $d$-electron gap coincides, including 
the case $V=0$) and with increasing $V$ an insulator-metal transition takes 
place at some critical value of $V=V_c$. In our preceding paper we reexamined 
these strong-coupling results by exact-diagonalization calculations and 
obtained fully different conclusions~\cite{Large_U}. In particular, we have 
found that for zero hybridization the gaps in the $f$ and $d$ electron density 
of states do not coincide, and almost all $f$ electron 
spectral weight is located outside the $d$ electron subbands.
For nonzero hybridization the $f$ and $d$ electron gaps coincide, 
for both the symmetric and unsymmetric case, but no insulator-metal 
transition driven by hybridization is observed in one as well as in two 
dimensions. In this contribution we extend numerical calculations 
to the opposite limit (the weak-coupling limit). From this point of view 
the paper represents the first systematic (exact) study of dynamical 
properties of the FKM in the weak-coupling limit. 
Here, the special attention is devoted to the behavior of the $f$ and 
$d$-electron density of states  with increasing $E_f$.  Such an analysis 
is very important since a parametrization of $E_f$ with applied 
pressure~$p$~\cite{SF} can, in principle, provide an interpretation of 
some experimental data, e.g., the behavior of the activation gap (the gap 
between the occupied and unoccupied states) with increasing $p$. 
In our previous paper~\cite{prb2} the problem of pressure dependence 
of the activation gap was analysed through the behavior of the energy gap 
in the $d$-electron spectrum of the FKM and many similarities with
the experimental data on the valence fluctuating compound SmB$_6$
were found. These similarities are however only qualitative, since 
in the correct analysis one should take into account also the behavior
of the $f$-electron spectral functions. This was done by Park and 
Hong~\cite{Park} and a nice correspondence of theoretical and
experimental results has been obtained. Unfortunately, these results 
have been obtained on a very small cluster (consisting of only eight 
sites) and thus cannot be considered as definite. Here we reexamine 
the behavior of the activation gap in the FKM for both $V=0$ and $V>0$.
At zero hybridization we are able to present results on relatively large 
clusters for both the $d$ and $f$ electron density of states, and thus our 
results can be extrapolated satisfactorily to the thermodynamic limit. 
For nonzero hybridization we were able to perform small-cluster
exact-diagonalization calculations on lattices only slightly larger 
than used by Park and Hong~\cite{Park} ($L\sim 12$), however a fundamental 
different behavior of the model is observed already on such small
clusters.            

\section{Results and discussion}

Let us start the discussion of our results with the case $V=0$.
In Fig.~1 we present numerical results for the $d$ and $f$ electron
density of states obtained for $U=0.6$ and several values of the $f$-level
energy $E_f$. Since the ground state configurations of the FKM in the weak 
coupling limit are well-known~\cite{prb2,epj} (the most homogeneous
(insulating) configurations for $|E_f| < E_{fc}\sim 1.34$, and the phase 
separated (metallic) configurations for $|E_f| \geq E_{fc}$),  
the $d$-electron density of states can be calculated directly 
from the single particle spectrum of the FKM model for $V=0$.
For the periodic configurations this can be done analytically
in the thermodynamic limit~\cite{Lyzwa}, and for an arbitrary 
$f$-electron concentration a numerical diagonalization
is possible on very large clusters ($L=64000$). Unlike this case,    
the $f$-electron density of states has to be calculated by exact 
diagonalization calculations (Lanczos method~\cite{Dag}), thereby
the cluster sizes are strongly limited ($L \leq 24$). To minimize 
the finite-size effects on the $f$-electron spectra we have
performed numerical calculations for several cluster sizes at each value 
of $E_f$. The typical behaviors are shown in Fig.~2 for two selected
values of $E_f$. It is seen that finite-size effects are small and thus 
already results obtained for $L=24$ can be used satisfactorily to 
represent the behavior of large systems. Although we have used in the 
numerical calculations the relatively large value of $\epsilon=0.01$  
for the resolution~\cite{note}, the formation 
of the gap in the $f$-electron density of states at the Fermi level
(in all examined cases the Fermi level is located between the first and
the second peak) is apparent. It is also apparent that the gaps in the
$d$ and $f$ electron spectra do not coincide. Moreover, one can see     
(in accordance with results obtained in the strong-coupling 
limit~\cite{Large_U}) that practically all $f$-electron spectral weight is 
located inside the principal $d$-electron gap (the gap at the Fermi
energy). With increasing $E_f$ both the principal gap as well as the 
$f$-electron spectrum shift to higher energies. Since the finite size 
effects on the $f$-electron spectra are negligible for the cluster sizes used 
in our numerical calculations ($L=24$), and even the $d$-electron spectra 
can be obtained exactly on much larger clusters, one can try to
construct the $E_f$ dependence of all relevant energy gaps. In particular,
we have calculated the $d$-electron gap $\Delta_d$ (the gap in the
$d$-electron spectrum at the Fermi energy), the $f$-electron gap $\Delta_f$ 
(the gap in the $f$-electron spectrum at the Fermi energy), and 
the $fd$-electron gap $\Delta_{fd}$ (the gap between the occupied
$f$ subband and the empty $d$ subband (the activation gap).
To obtain the $d$-electron gap $\Delta_d$ it is sufficient to know
the $f$-electron distribution that minimizes the ground-state energy
of the FKM for given $E_f$ and $U$. As mentioned above, the ground-state
configurations of the FKM in the weak-coupling limit are the most
homogeneous configurations for $|E_f| < E_{fc}$, and 
the phase separated configurations for $|E_f| \geq E_{fc}$. 
While the most homogeneous distribution of $f$ 
electrons can be easily generated for arbitrary $E_f(n_f)$, the 
$f$-electron distribution in the phase separated configuration 
has to be determined numerically. This distribution can be found
in principle exactly, as shown in~\cite{prb2} or approximately 
as done in~\cite{epj}. Here we adopt the latter method, since it
allows to treat several times larger clusters ($L\sim 300$) and still to 
keep the high accuracy of computations. The results of numerical
calculations for $\Delta_d$ obtained using this method on the cluster 
consisting of 240 sites are presented in Fig.~3. It is seen that $E_f$
dependence of the $d$ electron gap qualitatively mimics the pressure 
dependence of the activation gap in SmB$_6$~\cite{Cool} (we note a
parametrization of $E_f$ with pressure). With increasing pressure (the 
$f$-level position $E_f$) the gap decreases and vanishes discontinuously 
at some critical pressure $p_c$ ($E_{fc}$). The finite-size scaling
analysis that we have performed for a wide range of $L$ values
($L=40,60 \dots  240$) showed that $\Delta_d$ is practically independent
of $L$ for all $|E_f|< E_{fc}$. On the other hand the same analysis
performed for $|E_f| \geq E_{fc}$ revealed that the finite-size effects
are still present in this region, but a disappearance of $\Delta_d$
can be satisfactorily confirmed by extrapolation of results obtained 
for different $L$.   

To determine the $E_f$ dependence of the $f$-electron gap $\Delta_f$ ,
and the $fd$-electron gap $\Delta_{fd}$ we have performed exhaustive
numerical studies of the $f$-electron density of states for a wide 
range of $E_f$ values. In particular, we have calculated the $f$-electron
density of states for each $E_f$ from the interval [0,1.4] with 
the step $\Delta E_f=0.05$, and with the step $\Delta E_f=0.01$
for $E_f$ near the point of the insulator-metal transition.
To reveal the finite-size effects on $\Delta_{fd}$ and $\Delta_{f}$
the calculations have been done for several cluster-sizes 
$(L=12,16,20,24)$ at each $E_f$. It was found that finite-size effects 
on $\Delta_{f}$ and $\Delta_{fd}$ are negligible for $|E_f|<E_{fc}$, 
while small finite-size effects have been observed for $E_f>E_{fc}$. 
The resultant $E_f$ dependence of $\Delta_{f}$ and $\Delta_{fd}$ is 
presented in Fig.~3. It is seen that $\Delta_{fd}$ exhibits the same 
behavior as $\Delta_{d}$, of course with one exception and namely that 
$\Delta_{df}\sim \Delta_{d}/2$. Since $\Delta_{fd}$ is the 
gap between the occupied $(f)$ and unoccupied $(d)$ states (the activation 
gap) its behavior can be directly compared with the behavior
of the activation gap in SmB$_6$~\cite{Cool}. Although data for 
$\Delta_{fd}$ are more scattered, both activation gaps exhibit
qualitatively the same behavior. This indicates that the spinless
FKM, probably the simplest model of correlated electrons can, in principle,
provide the correct physics for describing properties of real materials.

Let us now examine what happens if the hybridization is switched on.
One could expect that just the hybridization will improve the accordance 
between the theoretical and experimental results for the activation gaps,
since the hybridization generally smears behaviors. This conjecture
support also results by Park and Hong~\cite{Park} obtained for the
FKM with hybridization on a very small cluster consisting of only
eight sites. Their results obtained for the $E_f$ dependence of the 
activation gap and the $d$-electron concentration $n_d$ are summarized 
in Fig.~4. It is seen a nice correspondence of theoretical and experimental
data for both the activation gap as well as the $d$-electron concentration. 
Unfortunately, these results have been obtained on a very small
cluster and so cannot be considered as definite. Even, the actual behavior
of the activation gap and the $d$-electron concentration $n_d$ on large 
lattices can be fully different from one presented in Fig.~4. Indeed,
a comparison of their results obtained for $n_d$ at $V=0$ with our results 
presented in Fig.~3b reveals fully different behavior of $n_d$. While 
$n_d$ calculated by Park and Hong~\cite{Park} for $L=8$ is constant
for all values of $E_f$ from $E_f=0.8$ to the insulator-metal transition
point $E_{fc}=1.34$, our results calculated on much larger clusters ($L=240$)
exhibit strong dependence on $E_f$. In the next we reexamine exactly the 
behavior of the FKM with hybridization on the cluster consisting of eight 
sites, as well as on clusters slightly larger ($L=10,12$). We present new
results that significantly improve results obtained by Park and 
Hong~\cite{Park}    

To show the hybridization effects on $\Delta_{d}$,$\Delta_{f}$ and 
$\Delta_{fd}$ we have calculated exactly the $f$ and $d$ electron density
of states for $V=0$ and $V=0.02$ on a small finite cluster consisting
of $L=8$ sites. We chose this cluster size to be compatible with 
results of Park and Hong~\cite{Park}. The resultant behaviors are 
presented in Fig.~5. The most prominent difference between the 
results obtained for $V=0$ and $V=0.02$ is that a nonzero spectral
weight appeared in the $d$-electron density of states at the Fermi 
level for finite hybridization. Of course, this fact will change
dramatically the picture discussed above for $V=0$. This is clearly
demonstrates in Fig.~6a, where the $E_f$ dependence of $\Delta_{d}$,
$\Delta_{f}$ and $\Delta_{fd}$ is displayed for $V=0.02$ and $L=8$.
It is seen that all gaps coincide for nonzero $V$, what strongly
contradicts to the case of $V=0$. For a comparison we have displayed
in Fig.~6a also the behavior of the single particle excitation 
energy defined as $\Delta_s=E_G(L+1)+E_G(L-1)-2E_G(L)$, where 
$E_G(N)$ is the ground state energy for $N$ electrons. As one could expect 
(on the base of the strong-coupling results~\cite{Large_U})
the single particle excitation energy $\Delta_s$ coincides with
$\Delta_{d}$, $\Delta_{f}$ and $\Delta_{fd}$ for nonzero hybridization.
This result is very important from the numerical point of view since 
the single particle excitation energy can be calculated easily by other
methods (that allow to treat much larger clusters, e.g., the  density matrix
renormalization group method (DMRG)), and so it can be used satisfactorily 
for describing conducting properties of the model. As one can see 
from Fig.~6a the conducting properties of the FKM with hybridization 
are described very inaccurate within the small-cluster 
exact-diagonalization calculations. In the region where the metallic 
phase has been identified for $V=0$, the single particle excitation energy
increases and it is not clear if this is a consequence of finite
hybridization or finite-size effects. Since calculations performed
for $L=10$ and $L=12$ revealed relatively large finite-size effects we have 
decided to use DMRG method to verify the actual behavior of $\Delta_s$.
In comparison to exact diagonalization calculations the DMRG method allows 
to treat several times larger clusters ($L\sim50$) and still to keep the high 
accuracy of computations. We typically keep up to 128 states per block, 
although in the numerically  more difficult cases, where the DMRG results 
converge slower, we keep up to 400 states. 
Truncation errors~\cite{White}, given by the sum of 
the density matrix eigenvalues of the discarded states, vary 
from $10^{-8}$ in the worse cases to zero in the best cases. 
The typical behavior of $\Delta_s$ is displayed in Fig.~6b for two
finite clusters of $L=32$ and $L=36$ sites.
Although the finite-size effects are still present for cluster
sizes treated by DMRG method the metallic region seems to be 
satisfactorily verified also for nonzero hybridization.
However, there is one important difference between the case 
$V=0$ and the case $V=0.02$. While for $V=0$ both $\Delta_d$        
and $\Delta_{fd}$ vanish discontinuously, they seem to vanish 
continuously for nonzero hybridization. This indicates that the 
spinless FKM behaves fully differently for zero and non-zero 
hybridization and fully different are also the corresponding 
pictures of insulator-metal transitions. Since rare-earth
compounds exhibit both types of insulator-metal transitions
(discontinuous as well as continuous) their different behavior
under the external pressure can be interpreted directly as 
a consequence of absence (presence) of hybridization in a given
material. According to this conjecture, e.g., SmB$_6$ that exhibits
a discontinuous pressure-induced insulator-metal transition
should be a material with zero hybridization between $d$ and 
$f$ states. Indeed, SmB$_6$ is the highest symmetry of the 
$O_h$ point group and the on-site hybridization between $d$ and $f$ 
states is forbidden by inversion symmetry~\cite{Foglio}.
This confirms that the spinless FKM, in spite of its relatively
simplicity, can yield the correct physics for a description of 
real materials, e.g. SmB$_6$. Besides of a qualitative correspondence
between the theoretical and experimental data for the activation gap,
we can state another example that supports this conclusion. 
In Fig.~7 we have displayed in detail the $d$ and  $f$ electron density
of states near the Fermi level for several values of $E_f$ 
($E_f=0,0.1,0.2$). For this set of $E_f$ values the ground state
is characterized by $n_f=0.5$ what models very well the situation 
in SmB$_6$ at low temperatures~\cite{prb2}. This electronic structure
consisting of two wide and two narrow subbands strongly mimics the 
electronic structure that we have proposed several years ago for 
a description of low temperature resistivity data of 
SmB$_6$~\cite{Batko}. The analysis of these data showed that they 
can be explained by introducing a fine structure (consisting of two 
narrow bands) into the principal gap, in accordance with our results
presented in Fig.~7. Of course, all these comparisons are only 
qualitative since our results have been obtained for the one-dimensional
system at $T=0$, while the real experimental systems are three dimensional
and measurements are done at finite temperatures. In future work,
we plan to perform a similar analysis in higher dimensions and $T \neq 0$.
Moreover, we also want to examine the influence of other factors 
(the electron-phonon interaction, the orbital dynamics, etc.) that have 
been neglected in this version of the model.

In summary, the $f$ and $d$ electron density of states of the 
spinless Falicov-Kimball model were studied in the weak-coupling limit 
by exact diagonalization calculations. The resultant behaviors were used 
to examine the $d$-electron gap $\Delta_{d}$, the $f$-electron gap 
$\Delta_{f}$, and the $fd$-electron gap $\Delta_{fd}$ as functions of 
the $f$-level energy $E_f$ and hybridization $V$. It was shown that the 
spinless Falicov-Kimball model behaves fully differently for zero and finite 
hybridization between $f$ and $d$ states. At zero hybridization the energy 
gaps do not coincide ($\Delta_{d}\neq \Delta_{f} \neq \Delta_{fd}$),
and the activation gap $\Delta_{fd}$ vanishes discontinuously
at some critical value of the $f$-level energy $E_{fc}$. 
In all examined cases practically all $f$-electron spectral 
weight is located outside the $d$-electron subbands.
On the other hand, at finite hybridization all energy gaps
coincide and vanish continuously at the insulator-metal transition 
point $E_f=E_{fc}$. The results obtained are compared with 
experimental behaviors observed in SmB$_6$.    

\vspace{0.5cm}
This work was supported by the Slovak Grant Agency VEGA
under Grant No. 2/7021/20 and the Science and Technology 
Assistance Agency under Grant APVT-51-021602.
Numerical results were obtained using
the PC-Farm of the Slovak Academy of Sciences.

\newpage
Figure Captions

\vspace{0.5cm}
Fig.~1. The $f$ and $d$ electron density of states of the FKM obtained for 
$V=0, U=0.6$ and several different values of~$E_f$. The results for the
$f$ electron density of states are plotted for $L=24$, while the results 
for the $d$ electron density of states are plotted for $L=64000$.
The corresponding $f$-electron concentrations are: $n_f=0.5$ for $E_f=0$,
$n_f=4/10$ for $E_f=0.5$, $n_f=1/3$ for $E_f=0.8$, and $n_f=1/4$ 
for $E_f=1.2$.

\vspace{0.5cm}
Fig.~2. The $f$ electron density of states of the FKM obtained for 
$V=0, U=0.6$ and two different values of $L$ and $E_f$. 

\vspace{0.5cm}
Fig.~3. The $d$-electron gap $\Delta_d$, the $f$-electron gap $\Delta_f$, 
the $fd$-electron gap $\Delta_{fd}$, and the $d$-electron concentration
$n_d$ as functions of the $f$-level energy $E_f$ calculated for $V=0$
and $U=0.6$. The solid lines are guides to the eye.  

\vspace{0.5cm}
Fig.~4. The activation gap $\Delta_{fd}$ and the $d$-electron concentration
$n_d$ as functions of the $f$-level energy $E_f$ calculated for $U=0.6$
and $L=8$~(Ref.~7).

\vspace{0.5cm}
Fig.~5. The $f$ and $d$ electron density of states of the FKM obtained for 
$U=0.6, E_f=1.2, L=8$ and two different values of~$V$. The arrow denotes
the most prominent difference between the results obtained for $V=0$
and $V=0.02$.

\vspace{0.5cm}
Fig.~6. a) The $d$-electron gap $\Delta_d$, the $f$-electron gap $\Delta_f$, 
the $fd$-electron gap $\Delta_{fd}$, and the single particle excitation
energy $\Delta_{s}$  as functions of the $f$-level energy $E_f$ calculated 
for $V=0.02, L=8$ and $U=0.6$. b) The single particle excitation energy 
$\Delta_{s}$ as a function of the $f$-level energy $E_f$ calculated 
by DMRG method for the same values of $V$ and $U$. The solid and dashed
lines are guides to the eye. 

\vspace{0.5cm}
Fig.~7. The $f$ and $d$ electron density of states of the FKM obtained for 
$V=0, U=0.6$ and several different values of~$E_f$. The results for the
$f$ electron density of states are plotted for $L=24$, while ones 
for the $d$ electron density of states are plotted for $L=64000$.
In all cases the $f$-electron concentration is equal to $n_f=0.5$.
To see clearly the $f$-electron gap, the $f$-electron density of 
states has been plotted with high resolution ($\epsilon=0.001$). 


\begin{thebibliography}{999}
\bibitem{note} $\epsilon$ is a small (real) number introduced in the 
calculation of spectral properties to shift the poles of the Green's 
function into the complex plane (see Ref.~14 for details).
\end{thebibliography}
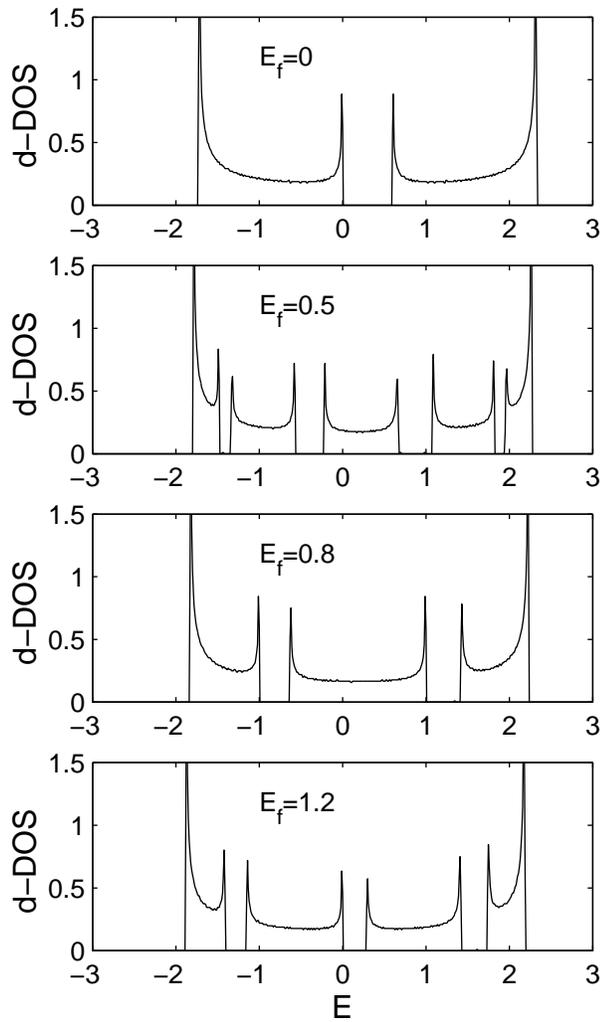
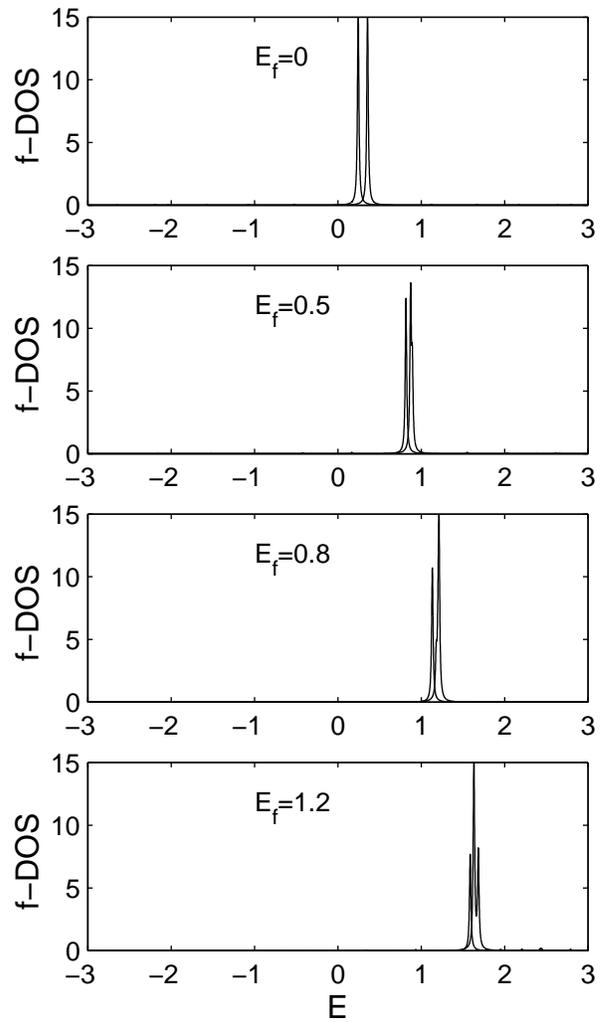
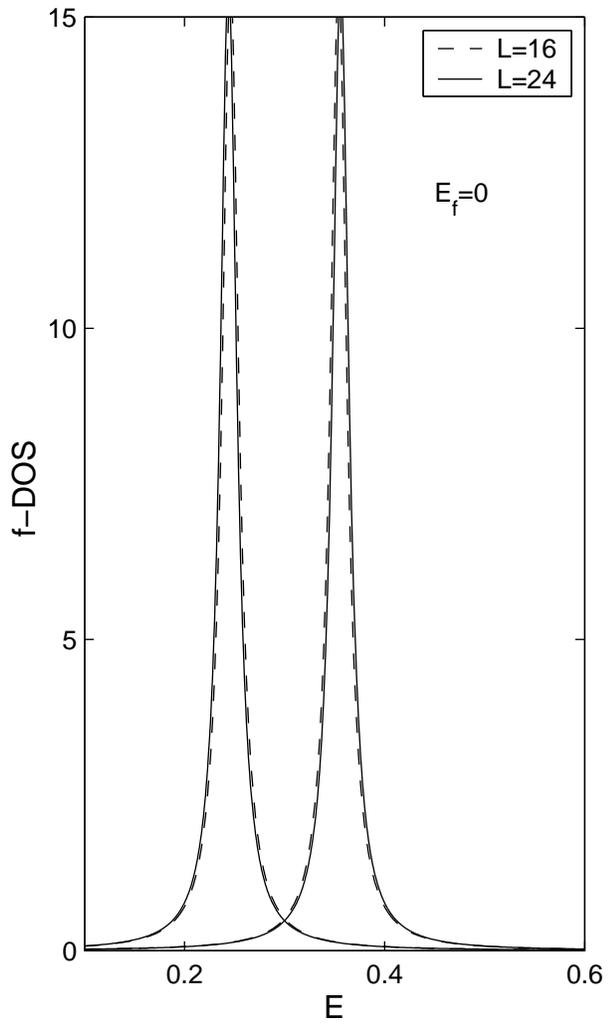
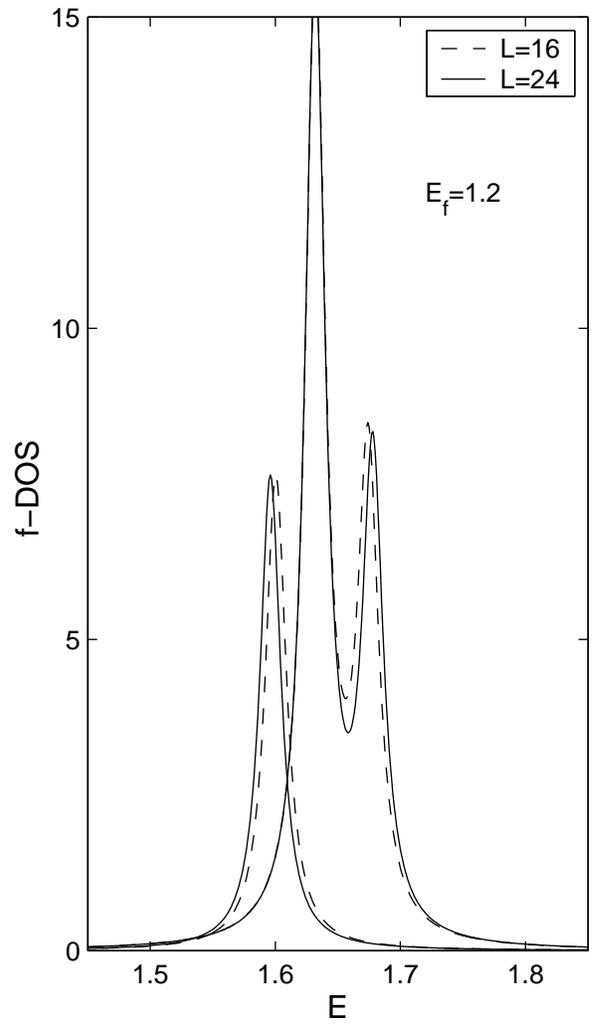
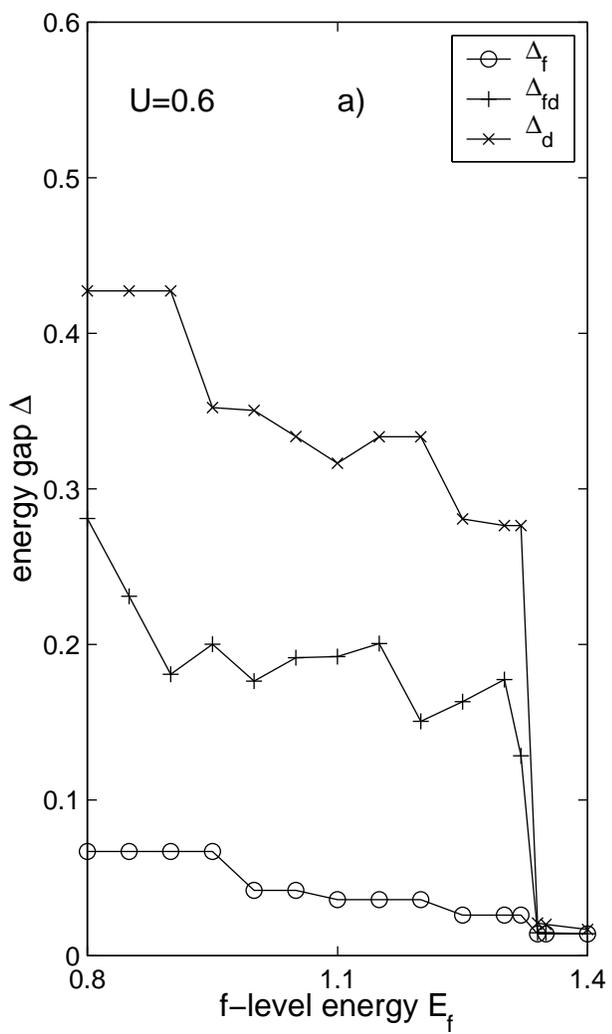
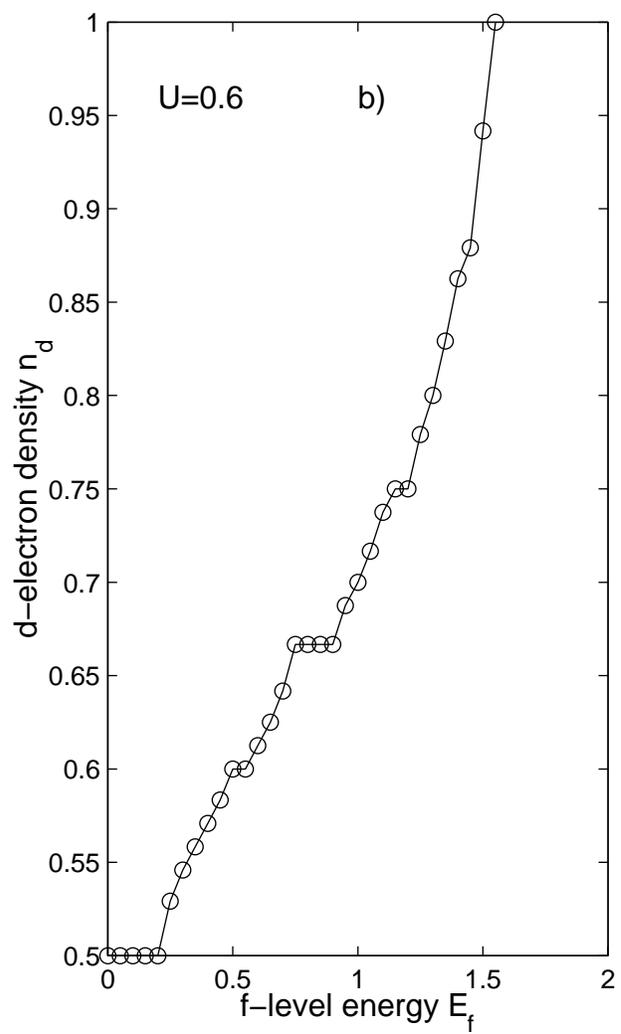
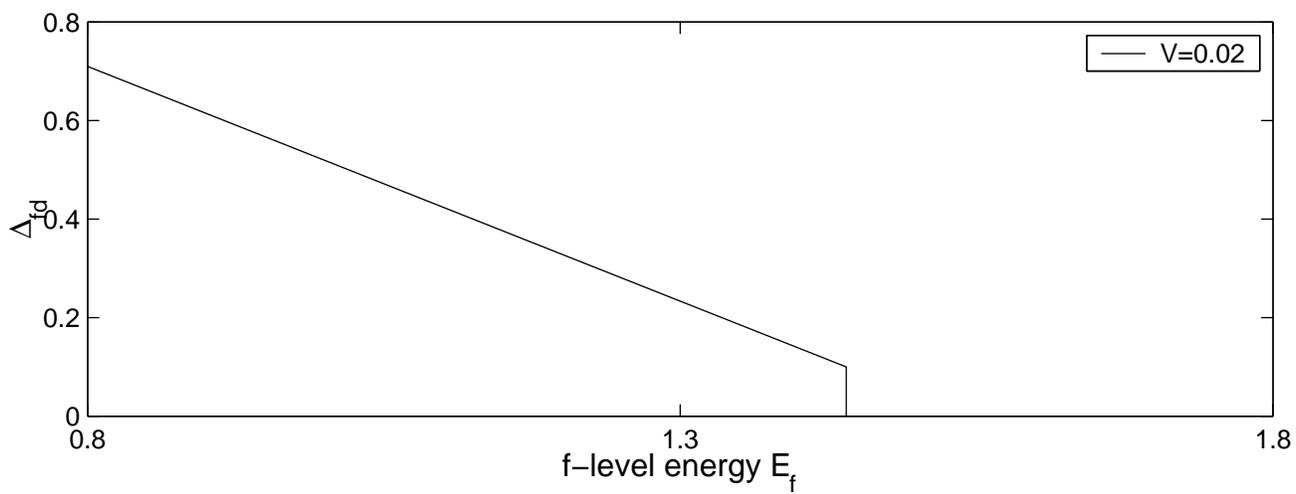
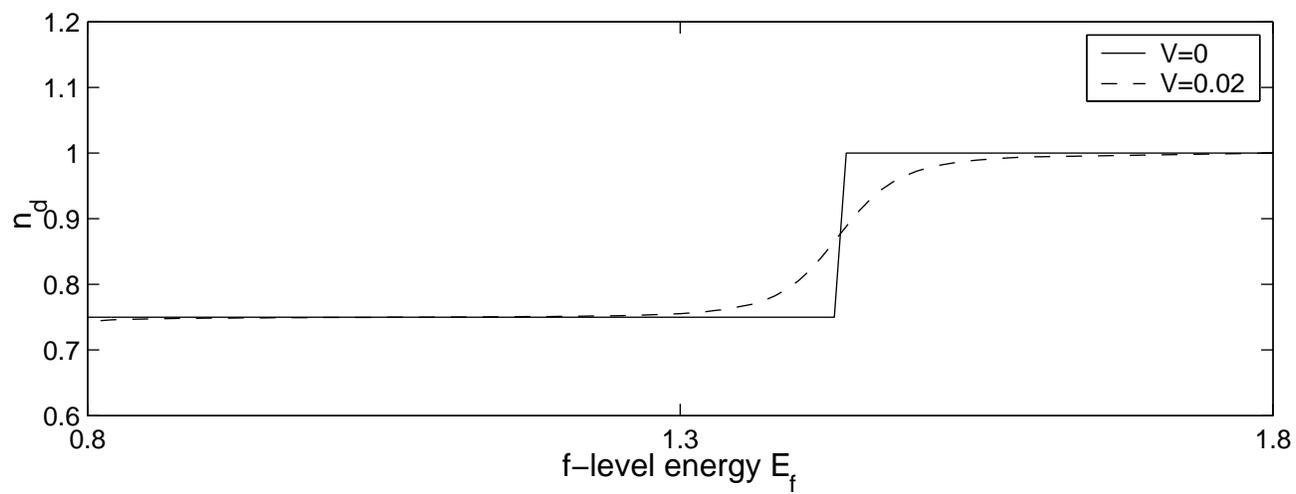
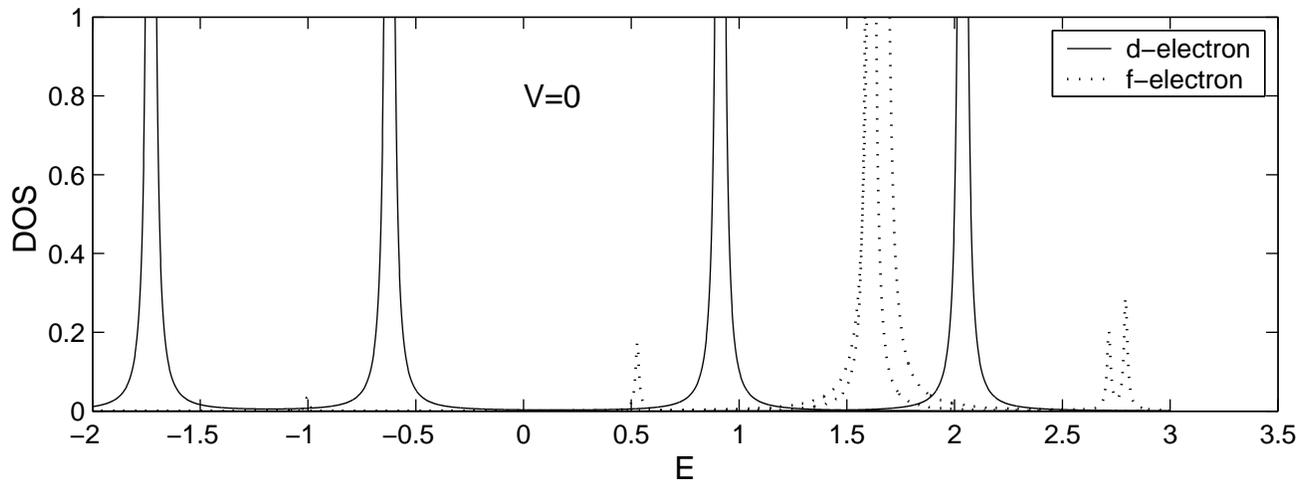
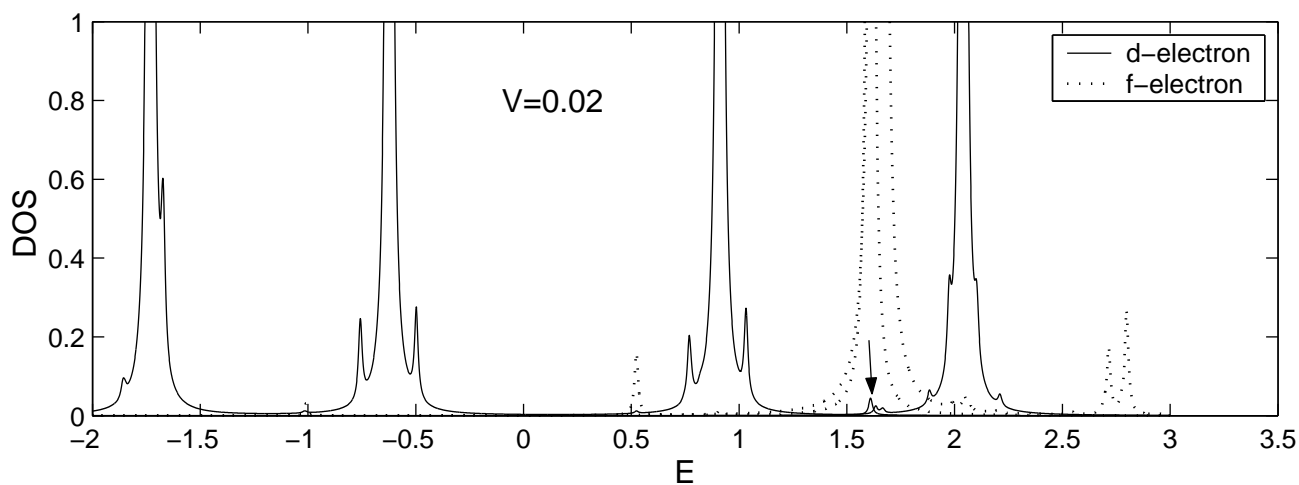
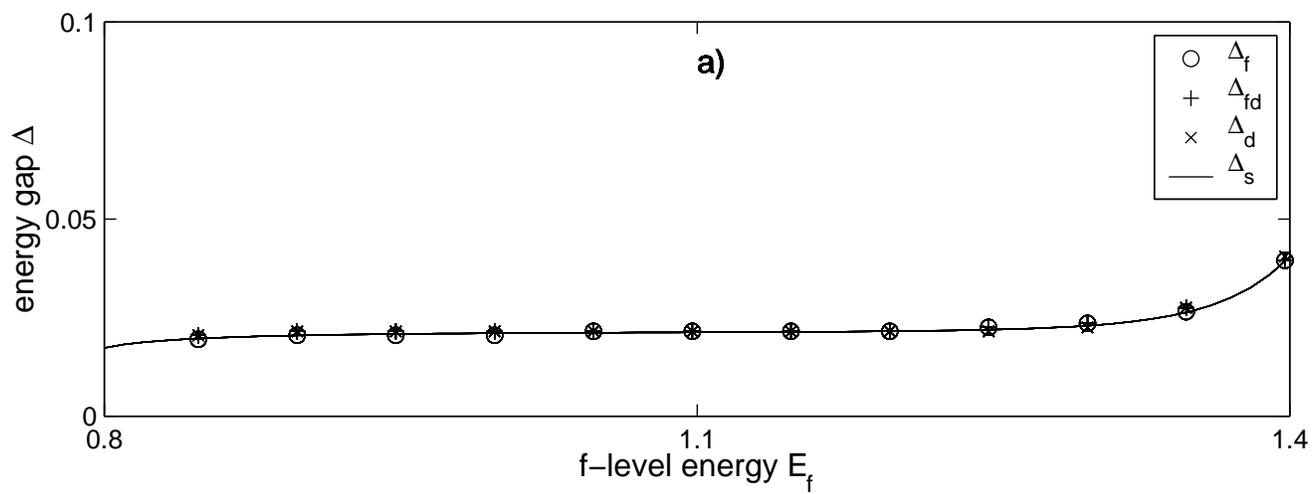
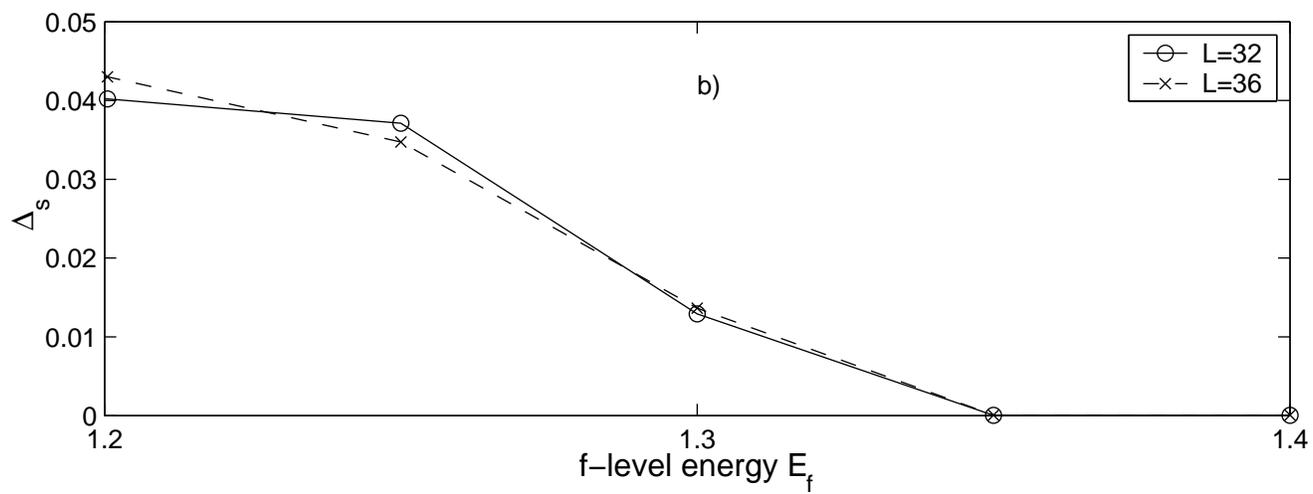
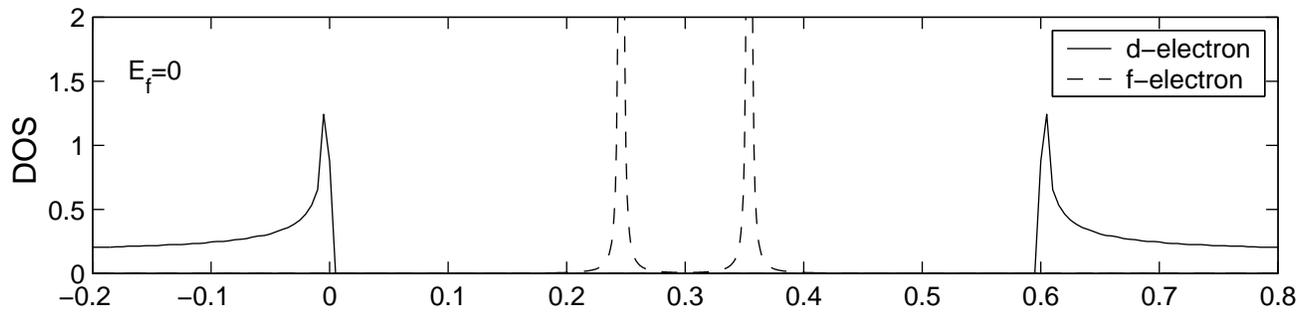
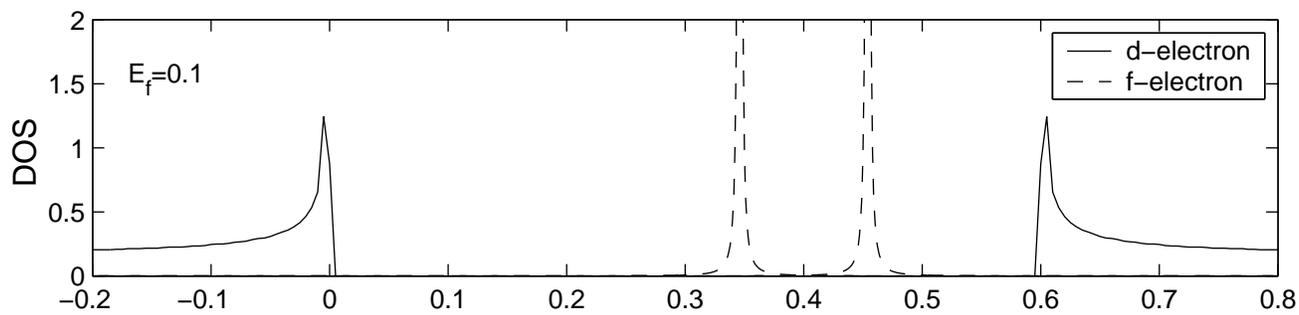
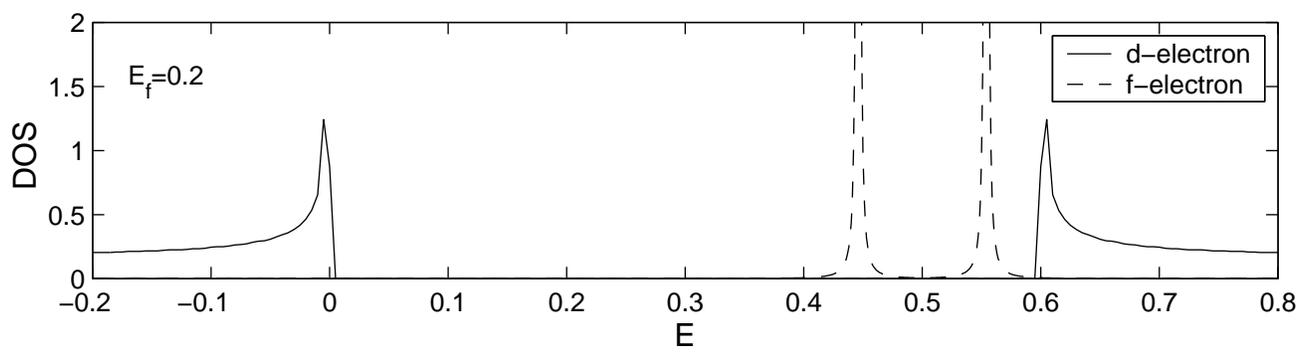
\end{document}